\documentclass[conference,a4paper]{IEEEtran}
\usepackage[utf8]{inputenc}
\usepackage[english,UKenglish]{babel}
\usepackage[T1]{fontenc}
\usepackage{microtype}
\usepackage[cmex10]{amsmath}
\usepackage{amssymb,amscd,upgreek,wasysym}
\usepackage{graphics}
\usepackage{graphicx}
\usepackage{epstopdf} % eps graphics
\usepackage{color}
\graphicspath{{./figures/}}

\DeclareMathOperator{\erf}{erf}
\DeclareMathOperator*{\amean}{mean}
\DeclareMathOperator*{\std}{std}
\DeclareMathOperator*{\var}{var}
\newcommand{\oneD}[1]{\underline{#1}}
\newcommand{\twoD}[1]{\underline{\underline{#1}}}

\begin{document}

\author{
\IEEEauthorblockN{Florian Wilde, Matthias Hiller and Michael Pehl\\}
\IEEEauthorblockA{Technische Universität München, Munich, Germany\\
 \{florian.wilde, matthias.hiller, m.pehl\}@tum.de} %\IEEEauthorrefmark{1}
}
\IEEEoverridecommandlockouts
\IEEEpubid{\makebox[\columnwidth]{978-1-4799-4833-8/14/\$31.00~\copyright2014 IEEE \hfill}\hspace{\columnsep}\makebox[\columnwidth]{ }}
\title{Statistic-Based Security Analysis\\of Ring Oscillator PUFs}
%\pagestyle{headings} 

%\authorrunning{masked}

\maketitle

\begin{abstract}
Ring oscillators (ROs) are a robust way to implement a physical unclonable function (PUF) into ASICs or FPGAs, but claims of predictability arose recently.
We describe why this likely results from not using adjacent ROs for pairwise comparison because of spatial patterns in mean frequency and correlation coefficients found by principal component analysis.
We show that the covariance is too small for our approach to estimate bits if adjacent ROs are compared.
Our assumption of normality for the inter-device distribution passes an Anderson-Darling test and we outline that devices with proximate serial numbers are not more similar than other devices.
\end{abstract}

\section{Introduction}
A physical unclonable function (PUF) utilises unavoidable manufacturing tolerances to derive a secret key unique to the device for secure identification, encryption, etc.
The use of Ring Oscillators (ROs) as PUFs is based on the assumption that these tolerances will impair the frequency of individual ROs in an independent and random manner so that the exact frequencies are unpredictable.
To derive a binary key from a set of ROs, it is preferable to use a differential evaluation, e.g. a pairwise comparison as in \cite{Mae12}, because the frequencies are also susceptible to environmental changes.
According to experimental data from an ASIC implementation \cite{Mae12}, RO-PUFs using this are almost ideally unique and by far the most reliable PUFs when using intra and inter Hamming distance \cite{Mae12} as measure for reliability and uniqueness, respectively.
In \cite{KKR+12}, using the same experimental dataset, the results regarding Hamming distance are verified, but the PUF output is said to be compressible by context tree weighting to 77\% of their original size, indicating significant redundancy and a lack of entropy.
This would open up a way to estimate bits from other ones, thus introducing a vulnerability.
Although \cite{Mae12} and \cite{KKR+12} use the same dataset, a substantial difference becomes evident on a closer look:
While in \cite{Mae12} adjacent ROs have been compared, in \cite{KKR+12} explicitly non-adjacent ROs have been compared. %, because the authors suspected adjacent ROs to be influenced by each other.
\textbf{Our contribution.}
We apply a selection of statistical methods on the data published by Maiti et al. in \cite{SES}, who recorded 100 samples of the frequencies of 512 ROs each implemented on 193 Xilinx Spartan (XC3S500E) FPGAs at room temperature.
The resulting statistical figures aid in the understanding of correlation processes important for the security of PUF realisations.
\textbf{Outline.}
Section~\ref{sec:distribution} analyses the inter-device distribution of the RO frequencies, followed by Section~\ref{sec:similarity} investigating similarity of devices and Section~\ref{sec:entropy} calculating entropy estimates.
In Section~\ref{sec:fitting} an approach to estimate bits via covariance matrices is taken and in Section~\ref{sec:pca} we apply a principal component analysis (PCA) to inspect correlations in detail before we finally conclude in Section~\ref{sec:conclude}.

\section{Nomenclature}\label{sec:nomenclature}
To reduce the influence of noise and environmental changes on our statistical analysis, we calculate the mean of the 100 readings from each individual RO provided in \cite{SES} and arrange them in a matrix $\twoD{F}$ whose elements are referred to as
\begin{equation}
F_{i,j} \qquad i \in \{0,\mathellipsis,511\} \quad j \in \{0,\mathellipsis,192\}
\end{equation}
where row number $i$ is the index of the RO and column number $j$ is the index of the FPGA.
To analyse dependencies of ROs within a device, we use the deviation of an RO's frequency from the mean frequency of the device it resides on as
\begin{equation}
D_{i,j} = F_{i,j} - \amean_{m=0}^{511} (F_{m,j})\label{equ:def_D}
\end{equation}
which form matrix $\twoD{D}$ of the same size as $\twoD{F}$.
$\amean_{m=n}^{o}(z_m)$ denotes the arithmetic mean of $\{z_n,\mathellipsis,z_o\}$ herein.
To derive a matrix of output bits $\twoD{R}$, we first group adjacent ROs disjunctively to ring oscillator pairs (ROPs) and calculate their frequency differences
\begin{equation}
B_{k,j} = F_{2k+1,j} - F_{2k,j} \qquad k \in \{0,\mathellipsis,255\}
\end{equation}
composing matrix $\twoD{B}$.
The elements of $\twoD{R}$ then are defined as
\begin{equation}
R_{k,j}=\begin{cases}1& B_{k,j}>0\\0& B_{k,j}\leq0\end{cases}
\end{equation}

\section{Determination of Distribution}\label{sec:distribution}
An important characteristic in a statistical investigation and imperative for Sections~\ref{sec:similarity}, \ref{sec:entropy}, \ref{sec:fitting} and \ref{sec:pca} is the identification of the probability distribution, here among the devices.
I.e. we test 512 respectively 256 random variables each with 193 samples.
Dependencies are not yet considered.
Due to the nature of the manufacturing tolerances and because they apply similarly to all ROs, we assume a normal distribution for their frequencies, which implies the same for deviations and differences.
To verify this assumption, we performed an Anderson-Darling test \cite{NIST_HB} for normality on the frequency and on the deviation of each RO and on the frequency difference of each ROP using
\begin{equation}\begin{split}
A^2 = - \sum_{n=0}^{N-1} \left( \frac{2n +1}{N} \left( \right.\right.\ln&(\Phi(Y_n))\\
 + \ln&(1-\Phi(Y_{N-1-n})) \left.\left.\right) + 1\right)
\end{split}\end{equation}
where $\Phi(\cdotp\!)$ is the normal cumulative distribution function \cite{B08}, $\oneD{Y}$ the scaled (divided by standard deviation), centered (mean subtracted) and sorted in ascending order sample vector, and $N$ the number of samples.
$Y_n$ is the $n^{\mathrm{th}}$ element of $\oneD{Y}$.

Using the sample size correction
\begin{equation}
A^{\ast 2} = A^2 \, \left( 1 + 4 N^{-1} + 25 N^{-2} \right)
\end{equation}
to fit the results to the tabulated limits in \cite{Steph74}, normality has then to be rejected with 1\% significance level if $A^{\ast 2}$ is greater than 1.047.
From Table~\ref{tbl:ni_results}, showing selected quantiles of the results for frequencies, deviations and differences, can be read that the majority of all three is well below the critical value.
Less than one percent of the ROs exceed it with their deviation which can be counted as outliers.
It is therefore justified to assume a normal distribution of the frequencies of each RO(P) among the devices -- i.e. within the rows of $\twoD{F}$, $\twoD{D}$ and $\twoD{B}$.

\begin{table}
\centering
\caption{Results of the Anderson-Darling test for normality}\label{tbl:ni_results}
\begin{tabular}{rcccc}
& \multicolumn{3}{c}{quantiles of $A^{\ast 2}$} & \\
data tested & 50\% & 90\% & 99\% & max $A^{\ast 2}$ \\
\hline
frequencies & 0.3990 & 0.5040 & 0.5996 & 0.6753 \\
deviations & 0.3572 & 0.6403 & 1.0384 & 1.3465 \\
differences & 0.3534 & 0.6595 & 0.9078 & 0.9973
\end{tabular}
\end{table}

\section{Similarity Analysis}\label{sec:similarity}
\begin{figure}
\input{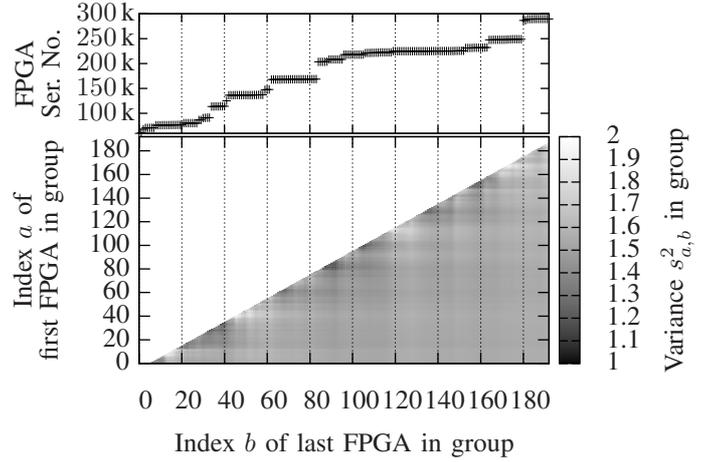}
\caption{Dissimilarity in groups of FPGAs together with the serial numbers of the FPGAs in the group}\label{fig:DF_grpvar}
\end{figure}

Important for the security of a PUF is whether the random process is white.
Which means here, that devices close regarding date of production (cf. upper part of \figurename~\ref{fig:DF_grpvar}) should not be more similar than devices who are distant.
To verify this, the mean empirical variance within groups of devices ranging from index $j=a$ to $b$ (inclusive) has been calculated by
\begin{equation}
s^2_{a,b} = \amean_{i=0}^{511}\left(\var_{j=a}^{b}(D_{i,j})\right)
\end{equation}
and plotted in \figurename~\ref{fig:DF_grpvar} for groups of at least five devices.
If the process would be non-white, the darker areas, i.e. smaller variance within the group, should coincide with the area where the plot of serial numbers of the FPGAs is flat.
This does not seem to be the case.
More formal, correlating $s^2_{a,b}$ with the difference between the serial numbers of FPGA $a$ and $b$ yields an absolute value of at most 0.21 for groups of ten devices, quickly decreasing for smaller or larger groups.
It is therefore generally not advantageous for an adversary to possess devices with serial numbers close to the one of the target device.

\section{Entropy Estimations}\label{sec:entropy}
An upper bound for the entropy of the PUF -- still assuming the bits to be independent -- can be derived from the sum of the bitwise entropies, which depend on the probability of each bit to be one (or zero).
This probability can either be derived from the bits itself
\begin{equation}
p_k = \Pr(R_{k,j}=1) = \amean_{j=0}^{192} (R_{k,j})\label{equ:bitbias_bin}
\end{equation}
or by modelling the random process and utilising the more detailed data available.
As mentioned in the previous section, the differences can be assumed normally distributed, yielding
\begin{equation}
p_k = \Pr(R_{k,j}=1) = \frac{1}{2} \left( 1 - \erf \left( \frac{\amean_{j=0}^{192}(B_{k,j})}{\sqrt{2} \std_{j=0}^{192}(B_{k,j})} \right) \right)\label{equ:bitbias_norm}
\end{equation}
where $\erf()$ is the Error-Function \cite{B08}.

\begin{figure}
\input{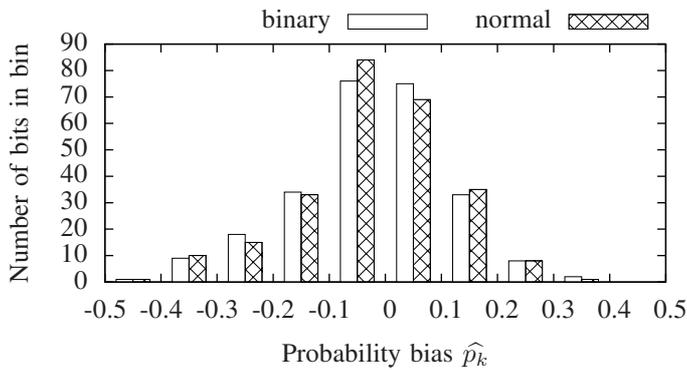}
\caption{Predictability of bits due to bias, binwidth 0.1}\label{fig:BF_bias_hist}
\end{figure}

\figurename~\ref{fig:BF_bias_hist} shows a histogram of the bias in probability
\begin{equation}
\widehat{p_k} = p_k - 0.5
\end{equation}
with $p_k$ according to \eqref{equ:bitbias_bin} and \eqref{equ:bitbias_norm}.
The entropy of the PUF output then follows by
\begin{equation}
H = - \sum_{k=0}^{255} \left( p_{k} \log_2(p_{k}) + (1-p_{k}) \log_2(1-p_{k}) \right)
\end{equation}
which gives 241.0 bits for the binary estimation and 241.3 bits for the normally distributed estimation, so about 94\% of the entropy achievable in 256 bits.
However, the practically usable entropy on a chip will be smaller, because some bits usually have to be masked due to reliability requirements.
Such masking is not done herein as the intra-instance distribution relevant for reliability is out of the scope of this paper.

\begin{figure}
\input{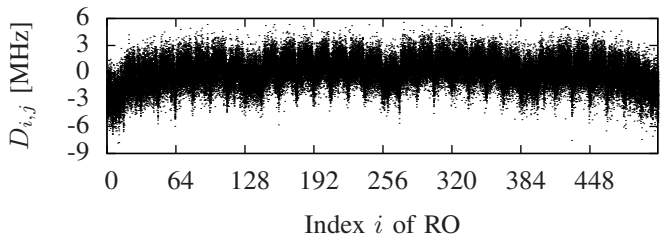}
\caption{Regular patterns in frequency deviations $D_{i,j}$}\label{fig:DF_scatter}
\end{figure}

An explanation for the bias can be derived from \figurename~\ref{fig:DF_scatter}, in which the elements of $\twoD{D}$ are visualised as dots.
While the variance seems to be similar for all ROs, the mean differs noticeable even between adjacent ROs.
Hence the amount of bias is merely defined by the difference of means and not also by differing variances.

As a consequence of these bias, not all keys are equally probable.
This opens up a way to get above linear proportion between the chance of finding the right key and the area of key-space searched.
One just needs to try the keys in descending order of probability.

Another method to calculate an upper bound for entropy \emph{considering dependencies} between instances is to compress the PUF outputs from various devices.
Using the reference implementation of context tree weighting (CTW) \cite{WST95} available from \cite{CTW} with default options and a binary file containing all elements of $\twoD{R}$ in column order, the highest compression rate achieved was $7.86237$ bits per byte, i.e. the size of the compressed data is still 98\% of their original size.
That this upper bound is worse than the priorly calculated suggests, that the compression algorithm was neither able to find inter-bit or inter-chip dependencies nor to fully utilise the lack of bitwise entropy.
Note that still both bounds rather support the claim of unpredictability in \cite{Mae12} than that of predictability in \cite{KKR+12}.

\begin{figure}
\input{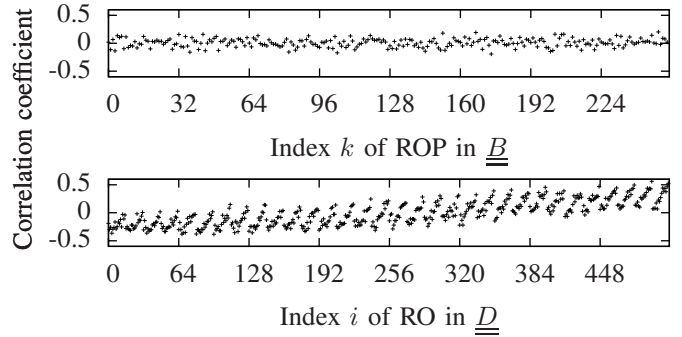}
\caption{Correlation with reference to RO(P) of highest index. Obviously distance depending for deviations $\twoD{D}$ but randomized for differences $\twoD{B}$}\label{fig:DBF_coco}
\end{figure}

The statement of minimal dependencies between the output bits is further supported by the comparably small (Pearson) correlation coefficients within $\twoD{B}$, of which a representative selection is plotted in \figurename~\ref{fig:DBF_coco}.
But this does not hold for the correlation coefficients of the deviations in $\twoD{D}$, a corresponding selection also plotted in \figurename~\ref{fig:DBF_coco}.
Their overall slope and the 32 smaller slopes indicate a dependency on the distance of the ROs.
This explains why the correlations vanish for $\twoD{B}$, because the pairwise comparison acts as a spatial high-pass filter.

\section{Covariance Fitting Approach}\label{sec:fitting}
Although the correlations in $\twoD{B}$ are small, they might still pose a threat by allowing to estimate the remaining bits after having guessed or otherwise become aware of some initial bits.
Our approach is based on the law of large numbers, which means as soon as the covariance matrix is calculated from a sufficiently large number of devices, an expanded covariance matrix including another device should differ only minimal.

If $\twoD{C}$ is the original covariance matrix obtained from training by $N\gg1$ devices with elements

\begin{equation}
C_{k,l} = \amean_{m=0}^{N-1} \left(\!\left(B_{k,m}\! - \amean_{n=0}^{N-1}(B_{k,n}) \right)\!\left(B_{l,m}\! - \amean_{n=0}^{N-1}(B_{l,n}) \right)\!\right)
\end{equation}
then $\twoD{\widehat{C}}$ is the expanded covariance matrix of $N+1$ devices with elements
\begin{equation}
\widehat{C}_{k,l} = \amean_{m=0}^{N} \left(\!\left(B_{k,m}\! - \amean_{n=0}^{N-1}(B_{k,n}) \right)\!\left(B_{l,m}\! - \amean_{n=0}^{N-1}(B_{l,n}) \right)\!\right)
\end{equation}
where the change in row mean values is assumed negligible to ease calculating the difference of both matrices.
The Euclidean norm of this difference will be the objective function to a non-linear least-squares minimisation algorithm
\begin{equation}
\min_{\oneD{B_N}} \| \twoD{\widehat{C}} - \twoD{C} \|^2_2
\end{equation}
The N\textsuperscript{th} column of $\twoD{B}$ $\oneD{B_N}$, corresponding to the estimated device, partially contains fixed values from guess or measurement.
The rest of the vector are free fitting variables to be adjusted by the minimisation algorithm to minimise the objective function.

\begin{figure}
\input{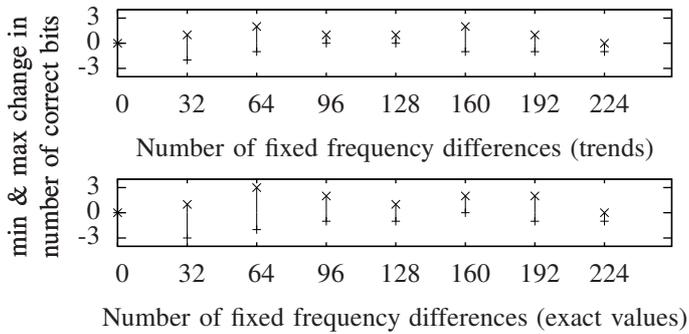}
\caption{Performance of minimisation approach via change in correctly estimated bits for eight randomly chosen FPGAs}\label{fig:BF_bitestim}
\end{figure}

This approach has been evaluated sequentially for eight FPGAs randomly chosen from the dataset and different amounts of fixed frequency differences.
The column of $\twoD{B}$ corresponding to the FPGA to be estimated was removed and $\twoD{C}$ calculated from the remaining 192 FPGAs.
As an attacker may not be able to measure any frequency difference but only to guess, the fixed values are set first to only their trend using $\pm1$ and afterwards to their exact values.
Starting point for the free variables was the respective mean.

The vertical scale of \figurename~\ref{fig:BF_bitestim} depicts the ability to estimate bits using the covariance fitting approach to be quite limited, it is even possible to get less correct bits than at starting point.

\begin{figure*}
\centering
\input{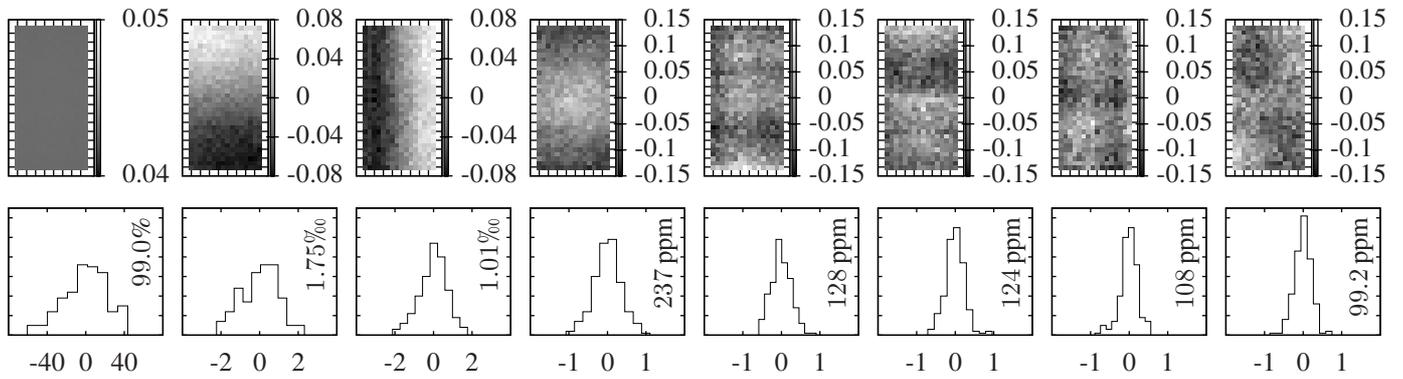}
\caption{Loadings (top), histograms of estimations (bottom) and described portion of variance (printed value) of the first eight principal components calculated from $\twoD{F}^T$. Note the different orders of magnitude for the printed values.}\label{fig:pca_vec_F}
\end{figure*}

\section{Principal Component Analysis}\label{sec:pca}
To take a closer look on the correlations in \figurename~\ref{fig:DBF_coco}, we use PCA, a statistical method to split correlated random variables into uncorrelated variables and vectors of correlation.
It can be efficiently calculated as a singular value decomposition (SVD), if the data is arranged in a matrix with each variable in a column, each observation in a row as well as scaled and centered such that the columns are zero-mean and unity-variance.

$\twoD{Y}$ contains the scaled data of $\twoD{F}^T$ and its SVD is
\begin{equation}
\twoD{Y} = \twoD{U} \, \twoD{D} \, \twoD{V}^T
\end{equation}
\textbullet\ Columns of $\twoD{V}$ are the orthonormal principal components, also called loadings, showing conjointly behaving variables\\
\textbullet\ Rows of $(\twoD{U}\,\twoD{D})$ are the estimations of the principal components (PCs) for each observation, i.e. device\\
\textbullet\ Non-zero elements of diagonal matrix $\twoD{D}$ are the square roots of the eigenvalues of the covariance matrix of $\twoD{Y}$, which are proportional to the amount of variance described by the PC

The first eight PCs, sorted left to right in descending order of described variance, are plotted in \figurename~\ref{fig:pca_vec_F} so as to reflect the location of the ROs on the devices.
Additionally, histograms of the respective estimations give an idea of how this PC is distributed among the devices.
The by far largest portion (about 99\%) of variance relates to the mean frequency of a device (Note: centering removes the mean among ROs with equal index), and is thus irrelevant as long as ROs from the same device are compared to derive an output bit.
The second and third largest portion of variance originate from linear frequency shifts along the y- and x-axis, respectively, of the device.
The following five PCs still show some kind of geometric pattern, though increasingly randomised.
Only the ninth and higher PCs look fully randomised and their histograms become increasingly narrow.
They are likely to contain the desired localised random effects responsible for the uniqueness of the PUF.

The spatial dependency through the first few PCs has major implications on the security of the PUF output.
If the ROs compared to derive an output bit are located far away from each other, especially the second and third PC will cause them to be similarly biased on a device even if there mean frequency among all devices (cf. \figurename~\ref{fig:DF_scatter}) is equal.
%Remember PCA uses centered data.
Although $\twoD{R}$ is in this paper derived from directly adjacent (along the x-axis) ROs, the number of ones in the key of a device is found to still correlate by 0.5 with the estimation of the third PC for the device.
This is presumably caused by the ROPs with small difference in mean, which either tend to zero or one depending on the sign of the third PC. %and by how much.
If spatial dependencies also apply to ASIC implementations, it explains why the entropy calculated in \cite{Mae12} using pairwise comparison of adjacent ROs is significantly larger than the one calculated in \cite{KKR+12} by comparing distant ROs, while both use the same raw data.

Because PCs describe descending amounts of variance, i.e. contain descending amount of information, they are well suited for dimension reduction and lossy compression.
In this case it is possible to correctly calculate 85\% of all bits using only the first 102 PCs.
This reduction in dimensions, though, is assumed to not ease e.g. a brute-force attack, as the resulting dimensions are no longer binary, but real numbers.
Based on the estimations, we assume that even when limiting and quantising them, it will not result in a smaller amount of keys to try.

\section{Conclusion}\label{sec:conclude}
In this work we justified the assumption of normality for the inter-device distribution of RO frequencies by an Anderson-Darling test and showed that the process is white.
Our entropy estimations argue that although single bits can be heavily biased, the overall entropy of RO-PUFs is satisfactory.
Remaining correlations despite comparing adjacent ROs could not be exploited by covariance fitting, but future work may by utilising more advanced algorithms.
The depicted level of unpredictability is found to critically depend on the proximity of ROs used for pairwise comparison, as there may exist significant spatial patterns both in mean and in correlation of frequencies.
Further reducing the potential of these patterns to impair the uniqueness of RO-PUFs is part of future work, too.

\paragraph*{Acknowledgements}
The authors would like to thank Markus Wamser for his helpful advice on PCA.
This work was partly funded by the German Federal Ministry of Education and Research in the project SMERCS, grant number 01DP12037A.

\bibliographystyle{IEEEtran}
\bibliography{IEEEabrv,statistical_analysis_of_ROs}

\end{document}